# Measuring Antimatter Gravity with Muonium

Daniel M. Kaplan,[*] Derrick C. Mancini,[†] Thomas J. Phillips, Thomas J. Roberts,[‡] Jeffrey Terry
*Illinois Institute of Technology, Chicago, IL 60616, USA*

Richard Gustafson
*University of Michigan, Ann Arbor, MI 48109 USA*

Klaus Kirch
*Paul Scherrer Institute and ETH Zürich, Switzerland*

ABSTRACT

We consider a measurement of the gravitational acceleration of antimatter, $\bar{g}$, using muonium. A monoenergetic, low-velocity, horizontal muonium beam will be formed from a surface-muon beam using a novel technique and directed at an atom interferometer. The measurement requires a precision three-grating interferometer: the first grating pair creates an interference pattern which is analyzed by scanning the third grating vertically using piezo actuators. State-of-the-art nanofabrication can produce the needed membrane grating structure in silicon nitride or ultrananoscrystalline diamond. With 100 nm grating pitch, a 10% measurement of $\bar{g}$ can be made using some months of surface-muon beam time. This will be the first gravitational measurement of leptonic matter, of 2nd-generation matter and, possibly, the first measurement of the gravitational acceleration of antimatter.

## INTRODUCTION

The gravitational acceleration of antimatter has never been directly measured.[1] A measurement could bear importantly on the formulation of a quantum theory of gravity and on our understanding of the early history and current configuration of the universe. It may be viewed as a test of General Relativity, or as a search for new, as-yet-unseen, forces, and is of great interest from either perspective.

General Relativity (GR), the accepted theory of gravity, predicts no difference between the gravitational behavior of antimatter and that of matter. This follows from the equivalence principle—a key assumption of GR—implying that the gravitational force on an object is independent of its composition. While well established experimentally, GR is fundamentally incompatible with quantum mechanics, and development of a quantum alternative has been a longstanding quest. Since all available experimental evidence on which to base a quantum theory of gravity concerns matter–matter interactions, matter–antimatter measurements could play a key role in this quest. Indeed, the most general candidate theories include the possibility that the force between matter and antimatter will be different—perhaps even of opposite sign—from that of matter on matter.[2]

---

[*] Email: kaplan@iit.edu
[†] Also at Argonne National Laboratory, Argonne, IL 60439, USA
[‡] Also at Muons, Inc.

Although most physicists expect the equivalence principle to hold for antimatter as well as for matter, theories in which this symmetry is maximally violated (i.e., in which antimatter "falls up") are attracting increasing interest[3] as potentially providing alternative explanations for three great mysteries of physics and cosmology: what happened to the antimatter, where is the dark matter, and what is the dark energy? But, theory aside, the force exerted by matter on antimatter is something that can be determined only by experiment.

## MEASURING ANTIMATTER GRAVITY

Experiments to measure the gravitational acceleration of antimatter have only recently become feasible. They require the production of significant quantities of low-energy antiprotons at particle accelerators, or of intense, cool beams of positive muons, and technology to use those beams to make neutral hydrogenic atoms. Others[4,5,6] are pursuing the first measurement using antihydrogen, an antiproton bound to a positron. We here discuss a measurement using muonium (M), an antimuon bound to an electron. M atoms will be made in a monoenergetic, low-velocity, horizontal beam directed at an atom interferometer.[7] The interferometer can measure, with few-picometer precision, the amount by which the atoms fall (or rise!). This could give a measurement of $\bar{g}$, the gravitational acceleration of antimatter on earth, to a precision of ~10% of $g$, using some months of beam time at Switzerland's Paul Scherrer Institute (PSI), or, possibly, using a new surface-muon beam at the Fermilab Booster or Main Injector. A $\bar{g}$ measurement with a precision of 1% or better should subsequently become possible using the first stage of the Project X linac at Fermilab.

### Antihydrogen Measurement

The ALPHA collaboration at the CERN Antiproton Decelerator has recently published the first limit on the gravitational acceleration of antihydrogen.[4] They form and trap antihydrogen atoms from low-energy antiprotons and positrons inside an octupole Penning–Ioffe trap. The ground state of hydrogen has four hyperfine levels, of which two (due to their magnetic moments) are low-field seeking, and two high-field seeking. Solenoidal (Penning trap) fields confine the charged constituents, allowing antihydrogen atoms to be formed; the octupole winding creates a magnetic-field minimum along the trap axis allowing the low-field-seeking states of neutral antihydrogen to be trapped for times exceeding 1000 s.[8] When the trapping magnetic fields are de-energized, the atoms drift to the beam-pipe walls and annihilate, and the surrounding detectors are used to determine whether a preponderance of annihilations take place on the upper or lower half of the beam pipe. Sensitivity is limited by the temperature of the trapped antihydrogen and the turnoff speed of the magnets, and the first limit on the gravitational/inertial mass ratio, $F$, is obtained as $-65 < F < 110$. The collaboration estimates that by laser-cooling the antihydrogen to 30 mK or below and increasing the magnetic-field decay time constant to 300 ms, they will achieve sensitivity adequate to distinguish $F = -1$ from $F = +1$. This is a challenging endeavor requiring (among other advances) the development of reliable lasers to drive the Lyman-alpha transition between the ground and first excited states.[9]

### Muonium Measurement

The muonium measurement is made using a precision three-grating interferometer (Fig. 1): the first grating pair creates an interference pattern which is analyzed by scanning the third grating vertically using piezo actuators.[10] Separating the gratings by one lifetime offers a convenient compromise between the magnitude of the deflection to be measured and the statistics of the sample. The $\tau = 2.2$ µs muonium lifetime[11] then implies a gravitational deflection $g\tau^2/2 = 25$ pm between the first and second gratings and an interferometric phase shift $\Phi = 2\pi\, g\tau^2/d \approx 0.003$ if $d$ = 100 nm grating pitch is used, with ≈14% M survival and ≈10% transmission to the detector. The necessary gratings can be fabricated using state-of-the-art nanolithography, including electron beam lithography and pattern transfer into a free-standing film by reactive ion etching. Detection is straightforward using the coincident positron-annihilation and electron signals to suppress background.[12] Measuring $\Phi$ to 10% requires grating fabrication fidelity, and interferometer stabilization and alignment, at the few-picometer level; this is within the current state of the art.[13] At the anticipated rate of $10^5$ M atoms/s, and taking decays and inefficiencies into account, the measurement precision is $0.3g$ per $\sqrt{n}$, where $n$ is the exposure time in days.[7]

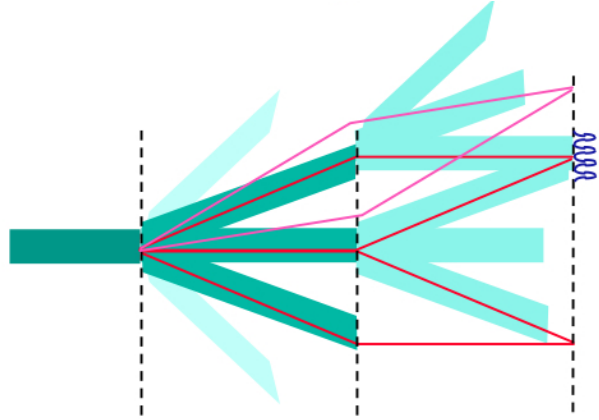

**Figure 1:** Principle of Mach Zehnder three-grating atom interferometer. The de Broglie waves due to each incident atom all contribute to the same interference pattern over a range of incident beam angles and positions. Each diffraction grating is a 50% open structure with a slit pitch of 100 nm. The assumed grating separation corresponds to one muon lifetime.

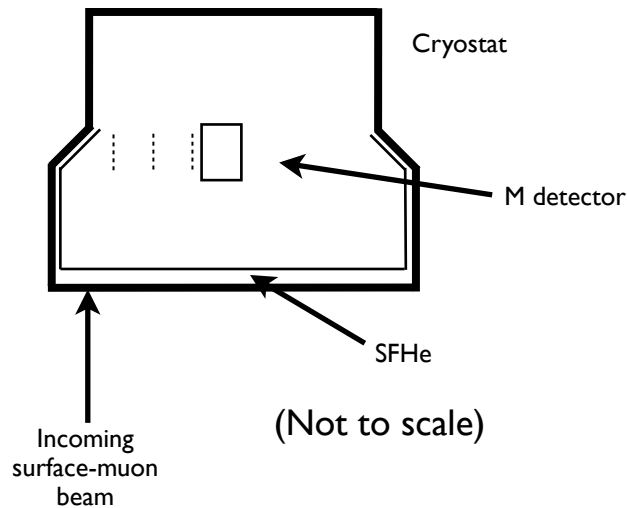

**Figure 2:** Concept sketch of muonium interferometer setup (many details omitted). A ≈micron-thick layer of SFHe (possibly with a small $^3$He admixture) stops the muon beam and forms muonium (M) which exits vertically and is reflected into the horizontal off of the thin SFHe film coating the cryostat interior.

Muonium is formed by stopping low-energy positive muon beams in matter. This is typically done in a powder, e.g., of $SiO_2$, providing stopping power along with voids that allow some of the produced M atoms to escape. This is an inherently inefficient technique: using the πE5 surface-muon channel at PSI, $8 \times 10^6$ μ$^+$/s at 26 MeV/c yielded an M rate of $1.6 \times 10^5$/s.[12] Moreover, it does not produce the narrow angular distribution needed for an interferometric gravity measurement. A new technique is proposed which can produce the needed beam. Positive muons entering from below are stopped in superfluid helium (SFHe) and M atoms emerge from the surface as a nearly monoenergetic, low-emittance, vertical beam with velocity of $6.3 \times 10^3$ m/s due to the chemical potential of M immersed in SFHe.[14] This beam can be reflected into the horizontal by a SFHe-coated surface at 45º. The short muonium lifetime requires a compact apparatus, with ≈1.4 cm between gratings, as sketched in Fig. 2.

## CONCLUSIONS

Whether the antihydrogen or the muonium measurement succeeds first is at this point uncertain, as both require new techniques with possibly a ~5–10-year development time. An unexpected result from either would imply a new understanding of gravity or the existence of new fundamental forces, not yet seen elsewhere, which may have played a key role in the evolution of the universe since the Big Bang.[2] Even confirmation of the predictions of GR would extend the equivalence principle to antimatter and constitute a classic test of that theory—one for the textbooks. Regardless of the antihydrogen result, the muonium measurement is uniquely sensitive to leptonic mass and to 2$^{nd}$-generation matter. Thus both measurements must be attempted: both are "obligatory homework assignments" from Mother Nature.

While not a part of the science case *per se*, it should also be noted that both antimatter and gravity have the potential to create great excitement among members of the general public—part of what we must do to maintain long-term funding for our field. This is an aspect that is well appreciated at CERN. It would be a great thing to be doing such research at Fermilab.

## ACKNOWLEDGMENTS

K.K. acknowledges fruitful discussions with D. Taqqu and L.M. Simons.

## REFERENCES

[1] M. Fischler, J. Lykken, T. Roberts, "Direct Observation Limits on Antimatter Gravitation," arXiv:0808.3929 [hep-th] (2008).
[2] M. Nieto and T. Goldman, "The Arguments Against Antigravity and the Gravitational Acceleration of Antimatter," Phys. Rep. **205**, No. 5 (1991) 221.
[3] See for example:
M. Villata, "On the nature of dark energy: the lattice Universe," Astrophys. Space Sci. **345** (2013) 1;
D. Hajdukovic, "Quantum vacuum and virtual gravitational dipoles: the solution to the dark energy problem?," Astrophys. Space Sci. **339** (2012) 1; "Quantum vacuum and dark matter," *ibid*. **337** (2012) 9;
A. Benoit-Lévy and G. Chardin, "Introducing the Dirac-Milne universe," Astron. & Astrophys. **537** (2012), A78;
M.Villata, "CPT symmetry and antimatter gravity in general relativity," EPL (Europhys. Lett.) **94** (2011) 20001;
M.Villata, "Gravitational interaction of antimatter," arXiv:1003.1635 [astro-ph.CO].
A. Benoit-Lévy, G. Chardin, "Do we live in a 'Dirac-Milne' universe?," arXiv:0903.2446 [astro-ph.CO];
L. Blanchet and A.L. Tiec, "Dipolar dark matter and dark energy," Phys. Rev. D **80**, 023524 (2009);
"Model of dark matter and dark energy based on gravitational polarization," *ibid*. **78**, 024031 (2008);

A. Burinskii, "The Dirac-Kerr-Newman electron," Gravitation and Cosmology **14**, 109 (2008);
L. Blanchet, "Gravitational polarization and the phenomenology of MOND," Class. Quant. Grav. **24**, 3529 (2007); "Dipolar particles in general relativity," *ibid*. 3541 (2007);
G. Chardin, "Gravitation, C, P and T symmetries and the Second Law," AIP Conf. Proc. **643** (2002) 385;
G. Chardin, "Motivations for antigravity in General Relativity," Hyp. Int., **109** (1997) 103
M. Kowitt, "Gravitational repulsion and Dirac antimatter," Int. J. Theor. Phys. **35** (1996) 605.

[4] C. Amole *et al*., "Description and first application of a new technique to measure the gravitational mass of antihydrogen," Nature Comm. **4**, 1785 (2013).

[5] A. Kellerbauer *et al*., "Proposed antimatter gravity measurement with an antihydrogen beam," Nucl. Instrum. Meth. B **266**, 351–356 (2008).

[6] G. Chardin *et al*., "Proposal to measure the gravitational behaviour of antihydrogen at rest," Report No. CERN-SPSC-2011-029/ SPSC-P-342 30/09/ 2011 (CERN, Meyrin, Switzerland, 2011).

[7] K. Kirch, "Testing Gravity with Muonium," arXiv:physics/0702143 [physics.atom-ph] (2007).

[8] G.B. Andresen *et al*., "Confinement of antihydrogen for 1,000 seconds," Nature Phys. **7**, 558 (2011).

[9] P.H. Donnan, M.C. Fujiwara, F. Robicheaux, "A proposal for laser cooling antihydrogen atoms," arXiv:1210.6103 [physics.atom-ph].

[10] See for example D.W.Keith, C.R. Ekstrom, Q.A. Turchette, and D. Pritchard, "An interferometer for atoms," Phys. Rev. Lett. **66**, 2693 (1991).

[11] A. Czarnecki, G.P. Lepage, W.J. Marciano, "Muonium decay," Phys. Rev. D **61** (2000) 073001.

[12] L. Willmann *et al*., "New Bounds from a Search for Muonium to Antimuonium Conversion," Phys. Rev. Lett. **82**, 49 (1998).

[13] For example, alignment and stabilization in LIGO are carried out to much higher precision than we need, as discussed in J. Abadie *et al*., "Calibration of the LIGO Gravitational Wave Detectors in the Fifth Science Run," Nucl. Instrum. Meth. A**624** (2010) 223.

[14] D. Taqqu, "Ultraslow Muonium for a Muon beam of ultra high quality," Phys. Procedia **17** (2011) 216–223. (Note that the beam formation technique we discuss is a simplified version of Taqqu's.)